\documentclass[aps, prd, twocolumn, nofootinbib,]{revtex4-2}

\usepackage[top=1in, bottom=1in, left=1in, right=1in]{geometry}

\usepackage{amsmath}
\usepackage{amsfonts}
\usepackage{wasysym}
\usepackage{graphicx}
\usepackage{comment}
\usepackage{tensor}

\def\filetype{pdf}

\def\path{}

\allowdisplaybreaks[1]

\begin{document}



\title{Dark matter admixed neutron stars}
\author{Ben Kain}
\affiliation{Department of Physics, College of the Holy Cross, Worcester, Massachusetts 01610 USA}

\begin{abstract}
\noindent Neutron stars could contain a mixture of ordinary nuclear matter and dark matter, such that dark matter could influence observable properties of the star, such as its mass and radius.  We study these dark matter admixed neutron stars for two choices of dark matter:\ a free Fermi gas and mirror dark matter.  In addition to solving the multi-fluid Tolman--Oppenheimer--Volkoff equations for static solutions and presenting mass-radius diagrams, we focus on two computations that are lacking in the literature.  The first is a rigorous determination of stability over the whole of parameter space, which we do using two different methods.  The first method is based on harmonic time-dependent perturbations to the static solutions and on solving for the radial oscillation frequency.  The second method, which is less well-known, conveniently makes use of unperturbed, static solutions only.  The second computation is of the radial oscillation frequency, for fundamental modes, over large swaths of parameter space.
\end{abstract} 

\maketitle


\section{Introduction}

If sufficient amounts of dark matter mix with the ordinary matter in a neutron star, then dark matter could influence measurable properties of the star.  This suggests the exciting possibility that neutron stars could act as laboratories for indirectly measuring dark matter properties.

An important question is how dark matter could become mixed with ordinary matter in a neutron star.  One well-studied possibility is through capture \cite{Goldman:1989nd, Kouvaris:2007ay, Bertone:2007ae, deLavallaz:2010wp, Kouvaris:2010vv, Brito:2015yga, Cermeno:2017xwb}.  If, in addition to gravitational interactions, dark matter has non-gravitational interactions with the ordinary matter in neutron stars, the extreme densities of neutron stars make them excellent targets.  Captured self-annihilating dark matter could potentially heat the star \cite{Kouvaris:2007ay, Bertone:2007ae, deLavallaz:2010wp}, while non-self-annihilating dark matter would accumulate.  For bosonic dark matter, this accumulation could lead to the formation of a small black hole which destroys the parent neutron star \cite{Goldman:1989nd, Bertone:2007ae, deLavallaz:2010wp, Kouvaris:2010vv, Brito:2015yga}, while for fermionic dark matter, degeneracy pressure is able to stabilize the star unless the dark matter particle mass exceeds $10^6$ GeV \cite{Gresham:2018rqo}.  Over the lifetime of a neutron star, the mass of the accumulated dark matter does not typically exceed $10^{-10}$ M$_\odot$ \cite{Goldman:1989nd, Kouvaris:2007ay, Ellis:2018bkr, Deliyergiyev:2019vti}, which has a negligibly small effect on the mass of the star.  

Another possibility for mixing is that the accumulation of dark matter occurs during stellar formation.  In \cite{Deliyergiyev:2019vti}, it is argued that a natural possibility is that the formation of a dark matter minihalo causes a neutron star to form from collapse.  (For other possibilities, see \cite{Nelson:2018xtr, Ellis:2018bkr}.)  Detailed studies of this process are lacking and would be interesting.  In this paper, we focus on bulk properties of the star, such as its mass and radius, and have this latter possibility for mixing in mind.

An appealing model for non-self-annihilating dark matter is asymmetric dark matter \cite{Kaplan:2009ag, Zurek:2013wia}, in which a conserved charge keeps dark matter from self-annihilating and an imbalance in the early universe between dark matter and anti-dark matter leads to the abundance of dark matter observed today.  A common description is a gas of Dirac fermions, possibly with self-interactions (see, for example, \cite{Gresham:2018rqo}).  Another description is as mirror dark matter \cite{Okun:2006eb}, which is motivated by the assumption that the Universe is parity symmetric.  The addition of new particles to restore parity to the Standard Model leads to mirror baryons as viable candidates for dark matter (see \cite{Khlopov1, Khlopov2, Khlopov3, Khlopov4, Sandin:2008db, Ciarcelluti:2010ji, Goldman:2011aa, Goldman:2013qla} and references therein).

Dark matter admixed neutron stars are two-fluid systems, in which the first fluid describes ordinary nuclear matter through an equation of state for a neutron star without dark matter and the second fluid describes dark matter.  Properties such as the mass and radius of the star are found by solving the multi-fluid Tolman--Oppenheimer--Volkoff (TOV) equations \cite{Kodama, Henriques:1989ar, Sandin:2008db}.  Null results from dark matter direct detection experiments \cite{Akerib:2017kat, Cui:2017nnn, Aprile:2018dbl} have placed stringent constraints on the dark matter-nucleon coupling strength.  From the perspective of the TOV equations, this is generally taken to mean that the dark matter-nucleon coupling strength is negligibly small \cite{Nelson:2018xtr, Gresham:2018rqo} and that dark matter admixed neutron stars are two-fluid systems in which the only inter-fluid interactions are gravitational.

Early work on dark matter admixed neutron stars was undertaken by Henriques, Liddle, and Moorhouse \cite{Henriques:1989ar, Henriques:1989ez, Henriques:1990xg} in their study of boson-fermion stars.  In these papers, the fermions were a free Fermi gas of neutrons and the scalar bosons could be interpreted as dark matter, though this was not explicitly stated.  Reference \cite{Henriques:1990xg} presented an underappreciated method for determining stability in two-fluid systems, which we make use of in Sec.\ \ref{sec:stability}.  Ciarcelluti and Sandin \cite{Sandin:2008db, Ciarcelluti:2010ji} used mirror baryons as dark matter.  Subsequent studies with mirror dark matter by Goldman et al.\ \cite{Goldman:2011aa, Goldman:2013qla} allowed for a mirror baryon mass smaller than the ordinary baryon mass.  A series of papers by Leung et al.\  \cite{Leung:2011zz, Leung:2012vea, Leung:2013pra} used a free Fermi gas as dark matter and studied situations in which dark matter forms either the core or the halo of the star.  As far as we are aware, Refs.\ \cite{Leung:2011zz, Leung:2012vea, Leung:2013pra} are the only papers that have computed radial oscillation frequencies for such systems using two-fluid methods.  A large number of studies have since followed \cite{Li:2012ii, Xiang:2013xwa, Li_2012, Tolos:2015qra, Mukhopadhyay:2015xhs, Panotopoulos:2017pgv, Panotopoulos:2017idn, Deliyergiyev:2019vti, Gresham:2018rqo, Nelson:2018xtr, Ellis:2018bkr, Bhat:2019tnz, DelPopolo:2020pzh, Zhang:2020dfi}, studying such things as self-interacting dark matter \cite{Li:2012ii, Xiang:2013xwa}, ordinary matter that includes hyperons \cite{Li:2012ii, DelPopolo:2020pzh} or strange quark matter \cite{Mukhopadhyay:2015xhs}, and a computation of the tidal deformability \cite{Nelson:2018xtr, Ellis:2018bkr, Zhang:2020dfi}.

In this paper, we study dark matter admixed neutron stars using two different models for asymmetric dark matter.  For the first model, we use a free Fermi gas.  Although self-interactions have been considered in a number of works and shown to lead to interesting effects (see, for example, \cite{Li:2012ii, Xiang:2013xwa, Gresham:2018rqo, Nelson:2018xtr, Deliyergiyev:2019vti}), for simplicity we do not include them.  For the second model, we use mirror dark matter, which was one of the first considerations in the study of dark matter admixed neutron stars \cite{Sandin:2008db}.

In addition to solving the TOV equations for static solutions, which gives the mass and radius of the star, we make a careful study of the stability of these solutions.  Rigorous determinations of stability with respect to small perturbations over large swaths of parameter space is lacking in the literature.  We present two different methods for determining stability.  The first method is to perturb the static solutions with harmonic perturbations and to solve for the squared radial oscillation frequency.  We do this using an approach developed in \cite{Kain:2020zjs}, which derived a system of pulsation equations for an arbitrary number of perfect fluids with only gravitational inter-fluid interactions and whose solution gives the squared radial oscillation frequency.  The second method we use was developed in \cite{Henriques:1990xg} and conveniently makes use of only unperturbed, static solutions.  Interestingly, we find regions of stable parameter space for which a naive analysis of the single-fluid equations of state would not have deemed stable.

Using the pulsation equations of \cite{Kain:2020zjs}, we also make a systematic determination of the radial oscillation frequencies for large swaths of stable parameter space.  This too appears to be lacking in the literature.  Although radial oscillation modes do not couple to gravitational waves, they are, in principle, observable \cite{Brillante:2014lwa} and the hope is that their study and detection can reveal details of the inner structure of the star.  We find interesting results here as well, in that the oscillation frequencies of dark matter admixed stars can be larger than the maximum possible frequencies of single-fluid stars made from the same equations of state.

In the next section, we review the multi-fluid TOV equations and the equations of state that we will be using.  In Sec.\ \ref{sec:stability}, we study stability, with some of the details given in  the Appendix.  In Sec.\ \ref{sec:MR}, we present mass-radius diagrams.  In Sec.\ \ref{sec:radial}, we compute radial oscillation frequencies.  We conclude in Sec.\ \ref{sec:conclusion}.


\section{Equations and equations of state}
\label{sec:equations}

Dark matter admixed neutron stars are solutions to the multi-fluid TOV equations \cite{Kodama, Henriques:1989ar, Sandin:2008db}.  The TOV equations follow from the Einstein field equations and the equations of motion in a spherically symmetric spacetime with static matter.  In writing equations, we use units such that $c = G=\hbar=1$.  The spherically symmetric metric may then be written as
\begin{equation} \label{metric}
ds^2 = - e^{\nu(r)} dt^2 + \frac{dr^2}{1-2 m(r)/r} + r^2 d\Omega^2,
\end{equation}
where $d\Omega^2 = d\theta^2 + \sin^2\theta \, d\phi^2$ and the metric function $m(r)$ gives the total mass inside a radius $r$.  For static solutions, the metric function $\nu(r)$ decouples and is not needed.  It is needed when determining radial oscillation frequencies and is discussed in the Appendix.

Null results from direct detection experiments for dark matter have placed stringent constraints on the dark matter-nucleon coupling strength \cite{Akerib:2017kat, Cui:2017nnn, Aprile:2018dbl}.  From the perspective of the TOV equations, this is generally taken to mean that any interaction between dark matter and the ordinary matter of the neutron star is negligibly small \cite{Nelson:2018xtr, Gresham:2018rqo} and that dark matter and ordinary matter can be modeled as separate fluids with only gravitational inter-fluid interactions.  The energy-momentum tensor, then, separates, $T^{\mu\nu} = \sum_i T^{\mu\nu}_i$, where the subscripted $i$ indicates the fluid (either ordinary or dark matter), and each energy-momentum tensor takes the perfect fluid form, 
\begin{equation} \label{static perfect fluid}
(T_i)\indices{^\mu_\nu} = \text{diag}( -\epsilon_i, p_i, p_i, p_i),
\end{equation}
where $\epsilon_i(r)$ and $p_i(r)$ are the fluid's energy density and pressure.  Each energy-momentum tensor is also conserved, $\nabla_\mu T^{\mu\nu}_i = 0$, which gives the equations of motion.  The TOV equations are then
\begin{equation} \label{TOV eqs}
\begin{split}
\frac{d m_i}{dr} &= 4\pi r^2 \epsilon_i
\\
\frac{d p_i}{dr} &= -\frac{4\pi r^3 p + m}{r^2(1-2 m/r)} (\epsilon_i + p_i),
\end{split}
\end{equation}
where the first equation follows from the Einstein field equations, the second from the equations of motion, $m = \sum_i m_i$, and $p = \sum_i p_i$.  In addition to the above equations, we shall need an equation for the number of particles inside a radius $r$, $\mathcal{N}_i(r)$, which is
\begin{equation} \label{N eq}
\frac{d \mathcal{N}_i}{dr} = \frac{ 4\pi r^2 n_i }{\sqrt{1 - 2 m/r}},
\end{equation}
where $n_i(r)$ is the number density.

With only gravitational inter-fluid interactions, the equations of state also separate, $\epsilon_i = \epsilon_i(p_i)$, where the energy density only depends on the pressure of the same fluid.  For ordinary matter, we use the analytical fit \cite{Haensel:2004nu} to the SLy equation of state \cite{Douchin:2001sv}.  SLy is a unified equation of state, obtained from a single effective nuclear Hamiltonian, allowing for a smooth transition between core and crusts of a neutron star.  The analytical fit further smooths the equation of state.  This level of smoothness is unnecessary for one of the methods we use to determine stability in the next section, but is helpful for the other method, which is also used to compute radial oscillation frequencies, because it requires taking derivatives of the equation of state.

It is worth noting that many equations of state in the literature for the ordinary matter of a neutron star, including \cite{Haensel:2004nu}, list the baryonic number density and not the number density for the fluid (i.e.\ the density of fluid elements).  But, as we will see, it is the fluid's number density that is needed for determining stability.  From the thermodynamic identity at zero temperature, $\epsilon + p - \mu n=0$, where $\mu = d\epsilon/dn$ is the chemical potential for the fluid, one finds that the number density for the fluid can be computed from the energy density and pressure,
\begin{equation} \label{n eq}
n \propto \exp \left( \int \frac{ d\epsilon}{\epsilon + p} \right),
\end{equation}
where the proportionality constant does not affect the determination of stability and therefore does have to be known.

As mentioned in the Introduction, we consider two possibilities for asymmetric dark matter.  For the first possibility, we use a simple and common description, modeling dark matter as a free Fermi gas.  The well-known equation of state and number density for a free Fermi gas are \cite{ShapiroBook, GlendenningBook}
\begin{align}
\epsilon &= 
\frac{1}{2\pi^2} 
\int_0^{k_{F}} dk \, k^2 \sqrt{k^2 + m_{f}^2}
\notag \\
&= \frac{1}{8\pi^2}
\Biggl[ k_{F} \sqrt{k_{F}^2 + m_{f}^2}( 2k_{F}^2 + m_{f}^2 )
\notag \\
&\qquad - m_{f}^4 \ln \left( \frac{k_{F} + \sqrt{k_{F}^2 + m_{f}^2}}{m_{f}} \right) \Biggr]
\notag \\
p &= 
\frac{1}{6\pi^2} \int_0^{k_{F}} dk \frac{k^4}
{\sqrt{k^2 + m_{f}^2}}
\notag\\
&= \frac{1}{24\pi^2}
\Biggl[ k_{F} \sqrt{k_{F}^2 + m_{f}^2}( 2k_{F}^2 - 3m_{f}^2 )
\notag \\
&\qquad + 3m_{f}^4 \ln \left( \frac{k_{F} + \sqrt{k_{F}^2 + m_{f}^2}}{m_{f}} \right) \Biggr]
\notag \\
n &= \frac{k_{F}^3}{3\pi^2},
\end{align}
where $m_f$ is the fermion mass and $k_F$ is the Fermi momentum.  The Fermi momentum is eliminated when forming $\epsilon = \epsilon(p)$ and $n = n(p)$, making the fermion mass the only free parameter.

For the second possibility, we consider mirror dark matter.  For this case, following \cite{Sandin:2008db, Ciarcelluti:2010ji, Goldman:2011aa, Goldman:2013qla}, we use the \textit{same} equation state for dark matter that we use for ordinary matter.

It is straightforward to show that the inner boundary conditions for the TOV equations in (\ref{TOV eqs}) and the particle number equation in (\ref{N eq}) are $m_i(r) = O(r^3)$, $p_i(r) = p_i^c + O(r^2)$, where $p_i^c$ is the central pressure for fluid $i$, and $N_i= O(r^3)$.  The central pressures uniquely identify solutions.  Upon specifying central pressures, the TOV and particle number equations may be integrated outward from some small $r$.  At some point during the integration, the pressure of one of the fluids will hit zero, $p_i(R_i) = 0$, marking the edge of fluid $i$ at $r = R_i$.  At this point, the integration is broken and restarted using the \textit{single}-fluid equations and the equation of state of the remaining fluid.  When the pressure of the remaining fluid hits zero, $p_j(R_j) = 0$, the edge of fluid $j$, as well as the edge of the star, is at $r = R_j$.  We then have for the total mass of the star, $M = \sum_i m_i (R_i)$, and for the total number of particles for fluid $i$, $N_i = \mathcal{N}_i(R_i)$.

Solutions to the TOV equations, when using the equations of state presented in this section, are examples of dark matter admixed neutron stars.  If $R_\text{dm} < R_\text{om}$, the star has a dark matter core, while if $R_\text{dm} > R_\text{om}$, it has a dark matter halo.  In the following, we display the parameter space of solutions using the central pressures $(p_\text{om}^c$, $p_\text{dm}^c)$, since they uniquely identify solutions.


\section{Stability:\ Critical curves}
\label{sec:stability}

Once a solution is found, an important question is whether it is stable with respect to small perturbations.  In systems with only a single fluid, this question is straightforward to answer, since it is well-known that the transition from stable to unstable occurs at the solution with the largest mass \cite{HarrisonBook, ShapiroBook, GlendenningBook}.  Since each solution is uniquely identified by a single quantity (the central pressure of the fluid), the static solution with the largest mass constitutes a single point in the parameter space of solutions and is called the critical point.  The critical points for single-fluid stars constructed with the SLy and free Fermi gas equations of state from the previous section are
\begin{equation} \label{critical points}
\begin{split}
(p^c_\text{SLy})_\text{crit} &= 
860.24
\,\,
\frac{\text{MeV}}{\text{fm}^3}
\\
(p^c_\text{Fermi})_\text{crit} &= 
291.20 \,
\left(\frac{m_f}{\text{1 GeV}} \right)^4
\,
\frac{\text{MeV}}{\text{fm}^3}.
\end{split}
\end{equation}

The situation is more complicated in two-fluid systems, such as dark matter admixed neutron stars.  Solutions are identified by two quantities (the central pressure of each fluid).  The transition from stable to unstable is then marked by a critical \textit{curve} in parameter space.  Determining the critical curve is not as straightforward as determining the critical point in the single-fluid case.  We offer two methods for its determination.

The first method is to perturb the static solutions with time-dependent harmonic perturbations, which depend on the radial oscillation frequency.  This leads to a system of pulsation equations, whose solutions give the squared radial oscillation frequency.  If the squared radial oscillation frequency is positive for the fundamental solution, the corresponding static solution is stable; otherwise it is unstable.  Such a system of pulsation equations was derived in \cite{Kain:2020zjs} for an arbitrary number of perfect fluids with only gravitational inter-fluid interactions and is reviewed in the Appendix (see also \cite{Comer:1999rs}).  

The second method was developed by Henriques, Liddle, and Moorhouse in their study of boson-fermion stars \cite{Henriques:1990xg}. The details are reviewed in the Appendix.  The conclusion of their analysis is that the critical curve is defined by
\begin{equation} \label{stability conditions}
\frac{d M}{d\mathbf{p}} = \frac{d N_\text{om}}{d\mathbf{p}} = \frac{d N_\text{dm}}{d\mathbf{p}} = 0,
\end{equation}
where $M$ and $N_i$ are the total mass and fluid number of a static solution and $\mathbf{p}$ is a vector in parameter space that is simultaneously tangent to the level curves of $M$ and $N_i$.  It can be shown that if two of the quantities in (\ref{stability conditions}) are zero, then the third is also  \cite{Henriques:1990xg, Jetzer:1990xa}.  We stress that $N_\text{om}$ is the total number of fluid elements and not the baryonic number.  This is the reason why the number density for the fluid must be known, which can be computed from the energy density and pressure using Eq.\ (\ref{n eq}).

In the original paper \cite{Henriques:1990xg}, the critical curve was found by plotting contour lines for $N_\text{om}$ and $N_\text{dm}$ and determining those points where the contour lines meet, but do not cross, so that their tangents are equal.  Such points give the critical curve.  An alternative procedure \cite{ValdezAlvarado:2012xc}, which is the one we use here, is to first compute contour lines of either $N_\text{om}$ or $N_\text{dm}$ in the two-fluid system.  Moving along a single contour line, we determine the point where $M$ is an extremum (in practice, we find that it is a maximum).  These points give the critical curve.

A benefit of the first method is that it can do more than just determine stability, since it can compute the radial oscillation frequency for an arbitrary static solution.  Its disadvantages are that it is time consuming to find a solution and it requires taking derivatives of the equation of state, which may be difficult to do if the equation of state is insufficiently smooth.  The second method does not suffer from either of these disadvantages, but can only determine stability.  We have confirmed numerically that both methods give the same answer, which gives confidence that the code we are using is working properly.  The figures presented in this section were made using the second method.  As far as we are aware, this is the first time that the second method has been applied to dark matter admixed neutron stars when a realistic equation of state is used for the ordinary matter.

\begin{figure}
\centering
\includegraphics[width=2.25in]{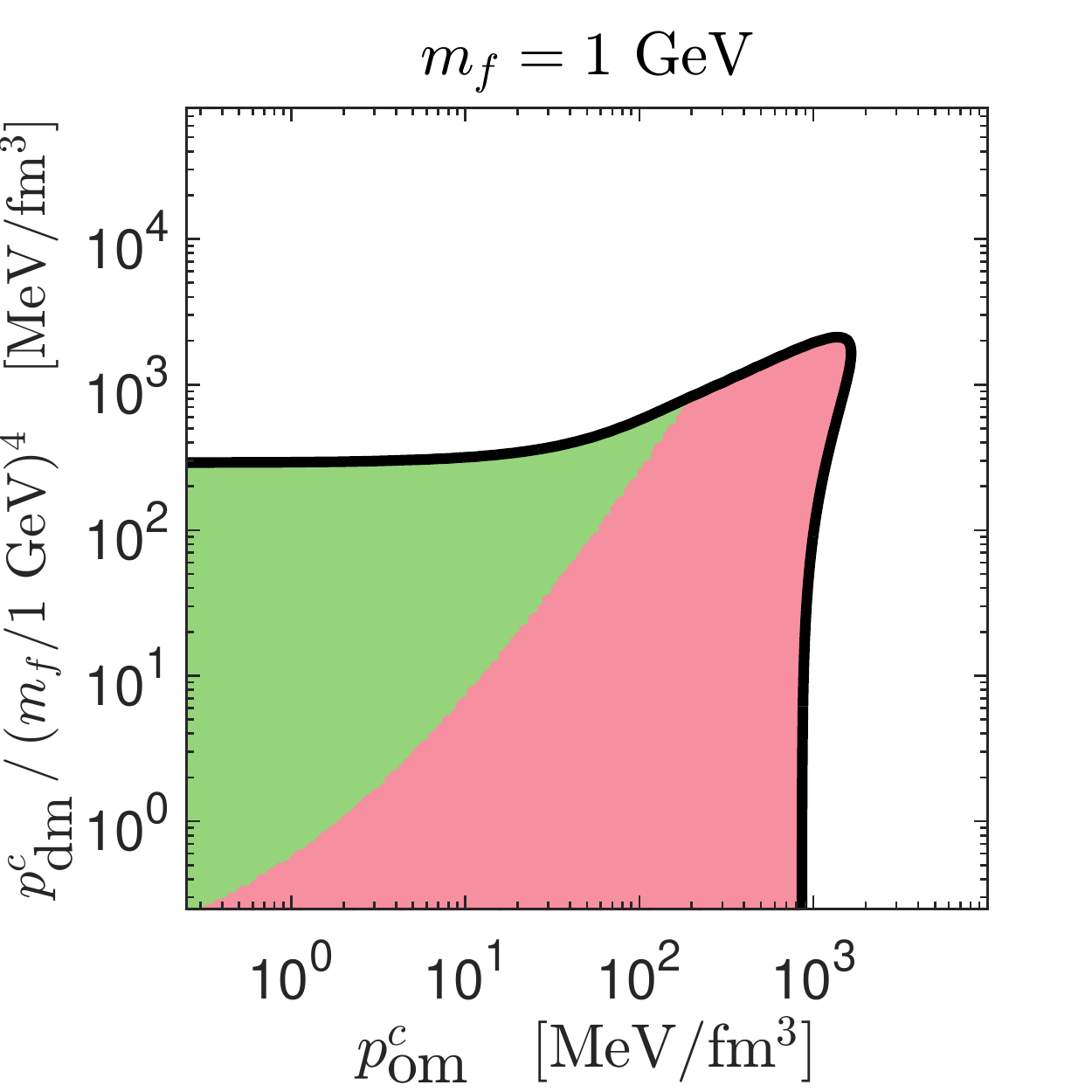}
\caption{The parameter space of static solutions is shown as a function of the central pressures $p^c_\text{om}$ and $p^c_\text{dm}$, where dark matter is taken to be a free Fermi gas with fermion mass $m_f = 1$ GeV.  Each point represents a static solution to the multi-fluid TOV equations.  The thick black line is the critical curve, separating stable static solutions from unstable ones.  Stable solutions are colored, with green indicating a static solution with a dark matter halo and red indicating a dark more core.}
\label{fig:stability m1}
\end{figure}

\begin{figure*}
\centering
\includegraphics[width=6.5in]{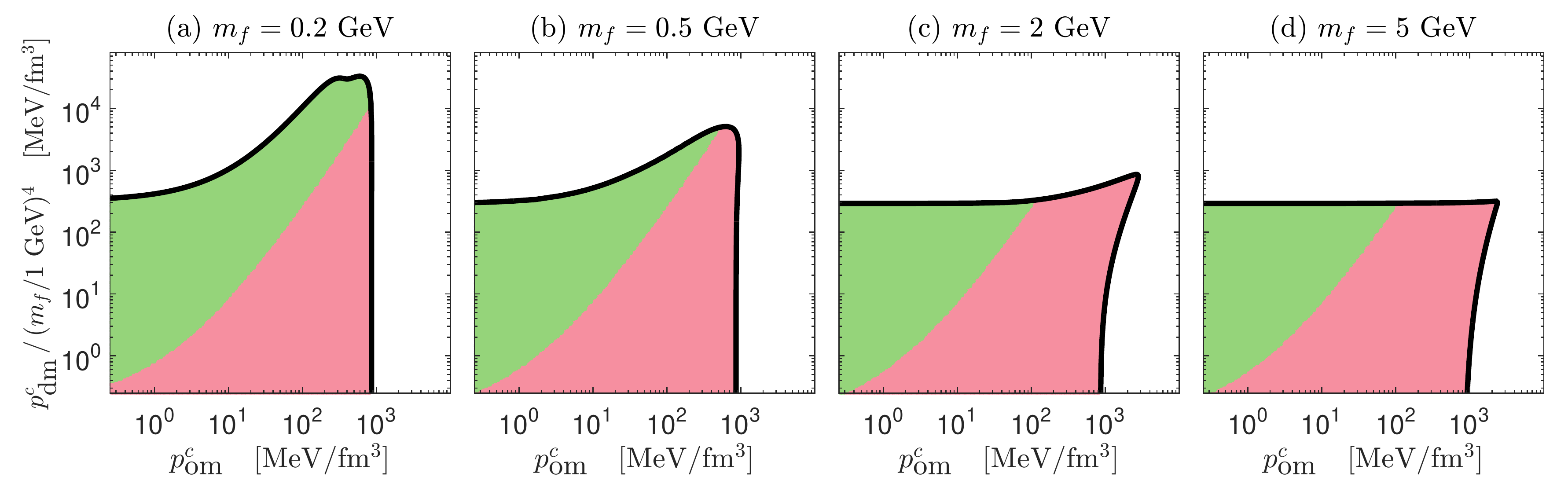}
\caption{The same as Fig.\ \ref{fig:stability m1}, except with fermion masses, $m_f$, as indicated above each plot. }
\label{fig:stability mneq1}
\end{figure*}

In Fig.\ \ref{fig:stability m1}, we show results for the free Fermi gas with fermion mass $m_f = 1$ GeV.  The thick black line is the critical curve.  The colored parameter space indicates stable static solutions, with green indicating static solutions with a dark matter halo and red a dark matter core.  For sufficiently small $p_\text{om}^c$ or $p_\text{dm}^c$, the critical curve is seen to agree with the single-fluid critical points in (\ref{critical points}).  This is expected, since if one of $p_\text{om}^c$ or $p_\text{dm}^c$ is small while the other is large, the fluid with the large central pressure dominates and we effectively have a single-fluid system.  Interestingly, there is a region of stable parameter space, in the upper-right corner, where $p_\text{om}^c$ and $p_\text{dm}^c$ are greater than their single-fluid critical values $(p_\text{om}^c)_\text{crit}$ and $(p_\text{dm}^c)_\text{crit}$ in (\ref{critical points}).

$m_f = 1$ GeV can be taken to approximate the transitional mass, where masses above and below this value lead to qualitatively different results.  The transitional mass is expected to be somewhere near the baryon mass of 938 MeV.  This is evident in Fig.\ \ref{fig:stability mneq1}, where we show critical curves for fermion masses above and below 1 GeV.  First, for $m_f$ below 1 GeV, we see from Figs.\ \ref{fig:stability mneq1}(a) and \ref{fig:stability mneq1}(b) that the critical curve moves toward extending beyond the single-fluid $(p_\text{dm}^c)_\text{crit}$, but not beyond the single-fluid $(p_\text{om}^c)_\text{crit}$.  This flips for $m_f$ above 1 GeV, where we see from Figs.\ \ref{fig:stability mneq1}(c) and  \ref{fig:stability mneq1}(d) that now the critical curve moves toward extending beyond the single-fluid $(p_\text{om}^c)_\text{crit}$, but not beyond the single-fluid $(p_\text{dm}^c)_\text{crit}$.  We do not compute the precise mass where this transition occurs, but simply take $m_f = 1$ GeV to approximate its value.

\begin{figure}
\centering
\includegraphics[width=2.25in]{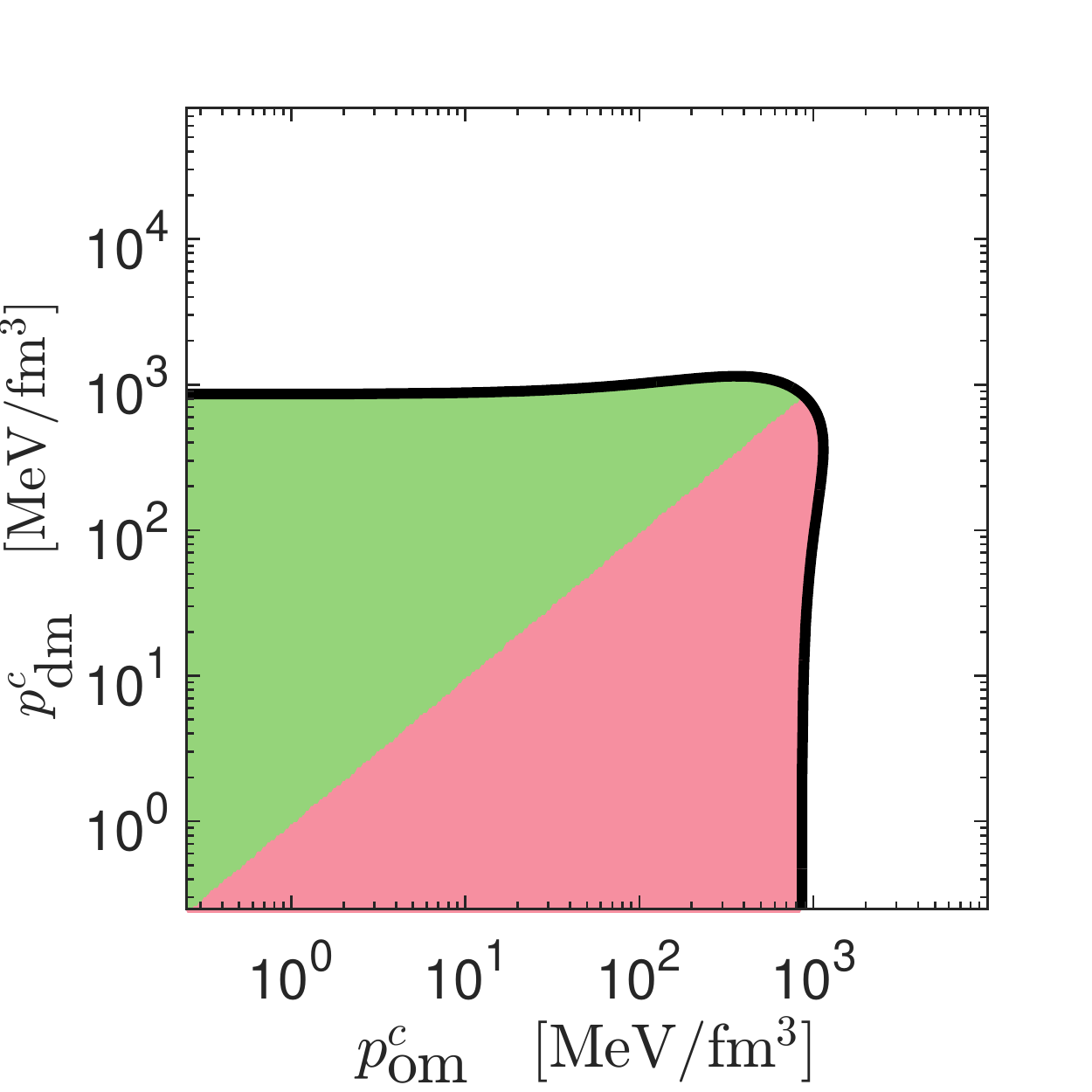}
\caption{The same as Fig.\ \ref{fig:stability m1}, except for mirror dark matter, in which dark matter has the same equation of state as ordinary matter.}
\label{fig:stability mirror}
\end{figure}

In Fig.\ \ref{fig:stability mirror} we show the critical curve for mirror dark matter.  Since the same equation of state is used for both ordinary and dark matter, both the critical curve and the line separating a dark matter core from a dark matter halo are symmetric in parameter space.  We find again that there is a stable region of parameter space where $p_\text{om}^c$ and $p_\text{dm}^c$ are greater than their single-fluid critical values, although the region is smaller than with the free Fermi gas.

In Sec.\ \ref{sec:radial}, when we look at radial oscillation frequencies, we will gain some insight as to why the critical curves extend past their single-fluid critical values and how this depends on fermion mass.


\section{Mass-radius diagrams}
\label{sec:MR}

Two of the most important properties to compute are the mass and radius of the star, since these properties are observable.  They are typically presented in mass-radius (MR) diagrams by plotting the mass as a function of radius for a given solution to the TOV equations.  In systems with only a single fluid, the mass-radius parameter space is a curve.  

Things are not as simple in two-fluid systems, such as dark matter admixed neutron stars, since the parameter space is larger.  In the literature, a single MR curve or a sequence of MR curves is often presented.  Unfortunately, this can give the false impression that there is an important property that connects two points on the same curve that does not connect two points on different curves.  While this is sometimes the case, commonly the property that connects two points on the same curve is arbitrary.  Another way to understand this is that any curve that slices through the parameter space in Figs.\ \ref{fig:stability m1}--\ref{fig:stability mirror} gives an MR curve, but there are infinitely many ways to slice through the parameter space.

\begin{figure}
\centering
\includegraphics[width=2.25in]{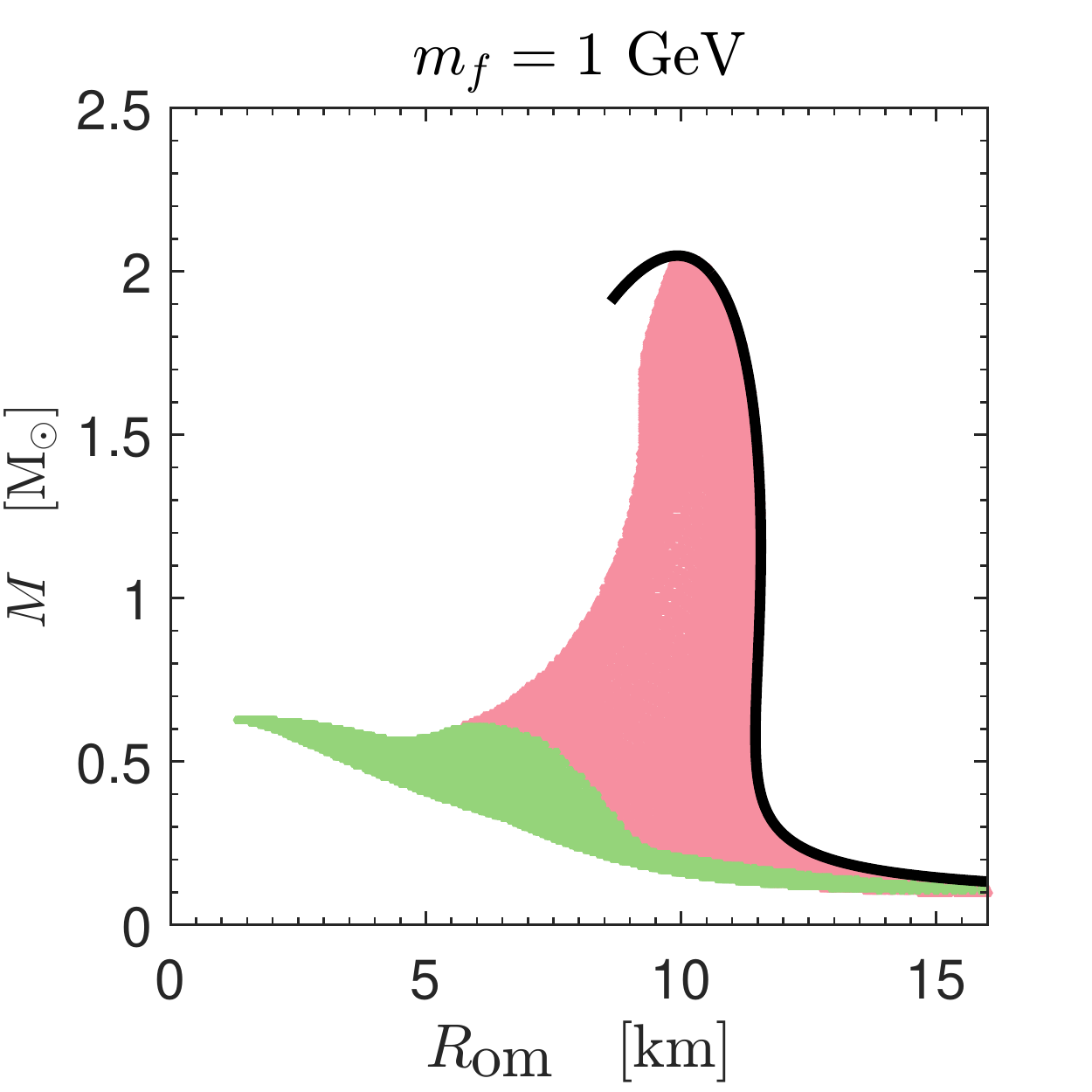}
\caption{The mass as a function of the visible radius, which is the radius of ordinary matter, for the stable static solutions shown in Fig.\ \ref{fig:stability m1} (dark matter is a free Fermi gas with fermion mass $m_f = 1$ GeV).  The thick black curve is for the single-fluid star with only ordinary matter, i.e.\ for a neutron star without dark matter.  The color scheme is the same as in Fig.\ \ref{fig:stability m1}, with green indicating a dark matter halo and red a dark matter core.}
\label{fig:MR m1}
\end{figure}

As an alternative, we present MR diagrams that show mass-radius relations for the entire stable parameter space shown in Figs.\ \ref{fig:stability m1}--\ref{fig:stability mirror}.  In Fig.\ \ref{fig:MR m1}, we display the MR diagram for the free Fermi gas with fermion mass $m_f = 1$ GeV.  We plot the total mass of the system versus the visible radius, which is the radius of  ordinary matter.  The thick black line gives the MR curve for the single-fluid star with the SLy equation of state, i.e.\ for a neutron star without dark matter.  The color scheme is the same as in Figs.\ \ref{fig:stability m1}--\ref{fig:stability mirror}, with green indicating a dark matter halo and red indicating a dark matter core.  For $m_f = 1$ GeV, we can see that much of the MR parameter space is taken up by a dark matter core.  Further, the inclusion of dark matter does not allow the mass or radius to extend past the values for a neutron star without dark matter.  In this sense, the inclusion of dark matter leads to a decrease in the mass and radius of the star \cite{Sandin:2008db}.

We mention again that all points plotted in Fig.\ \ref{fig:MR m1} are stable, as determined using the methods of the previous section.  Those points underneath the thick black line correspond to static solutions in which ordinary matter is dominating and we effectively have a single-fluid system.  As expected in such a case, stable static solutions do not extend past the peak of the thick black line.

\begin{figure*}
\centering
\includegraphics[width=6.5in]{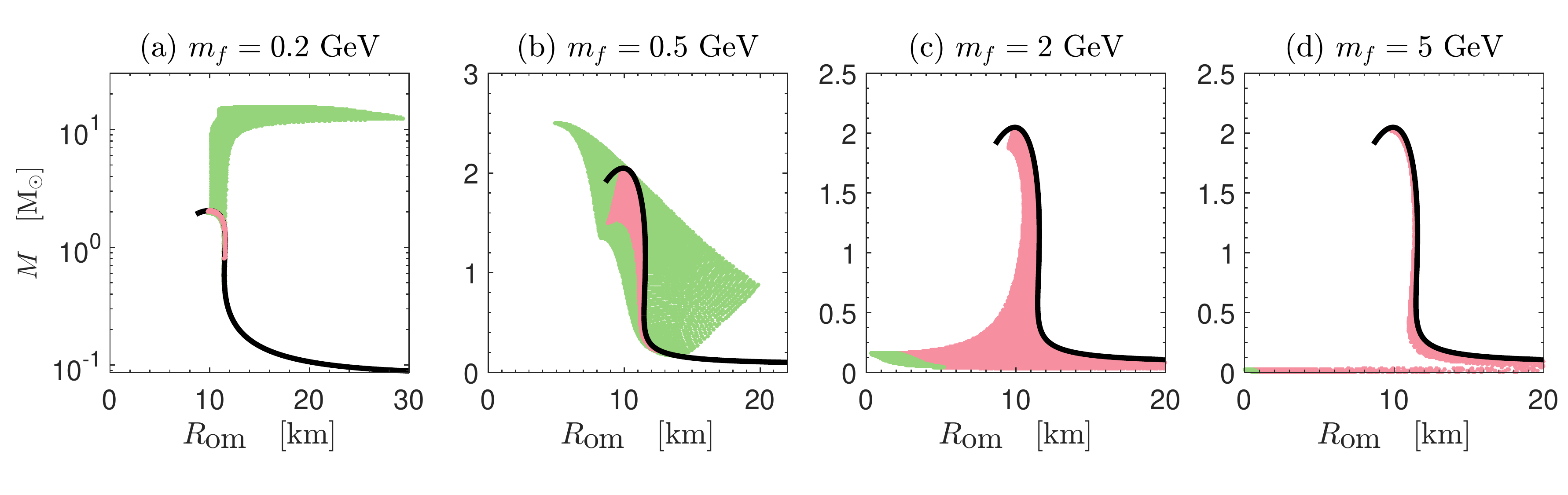}
\caption{The same as Fig.\ \ref{fig:MR m1}, except for the stable static solutions shown in Fig.\ \ref{fig:stability mneq1} (dark matter is a free Fermi gas with fermion masses, $m_f$, as indicated above each plot).  In (a), note that the vertical axis has a log scale and that the red region is plotted above the black curve so that it is visible.  In (b) note that the entire red region for a dark matter core overlaps green for a dark matter halo.}
\label{fig:MR mneq1}
\end{figure*}

We previously mentioned that $m_f = 1$ GeV is a transitional mass, in that larger and smaller masses give qualitatively different results.  This was evident in the previous section with stability and is also evident with MR diagrams.  In Fig.\ \ref{fig:MR mneq1}, we show MR diagrams for fermion masses above and below 1 GeV.  Beginning with Fig.\ \ref{fig:MR mneq1}(b), we see that as the fermion mass is lowered, the parameter space with a dark matter halo expands and the total mass of the system is able to increase past the values for a neutron star without dark matter.  This continues in Fig.\ \ref{fig:MR mneq1}(a) with an even smaller fermion mass.  We conclude that by decreasing the fermion mass, the total mass of the system can increase \cite{Goldman:2011aa} and dark matter generally encompasses ordinary matter, forming a halo.  What is happening is that as the fermion mass is decreased, which requires decreasing the dark matter central pressure to retain stability, dark matter has an increasingly weaker effect on ordinary matter.  In other words, as the fermion mass is decreased, ordinary matter begins acting as if it is a single fluid given by the black curve.  Not shown is that when dark matter forms a halo, the smaller fermion mass allows dark matter to extend farther out.  With dark matter extending farther out, more dark matter particles can stably exist, which can compensate for the smaller fermion mass and increase the total mass of the system.  On the other hand, when dark matter forms a core, there are not enough dark matter particles to compensate for the smaller fermion mass, which is why the red region is squeezing close to the black curve.

Now consider when the fermion mass is increased above 1 GeV.  We can see in Figs.\ \ref{fig:MR mneq1}(c) and (d) that as the fermion mass is increased, the parameter space becomes increasingly dominated by a dark matter core.  Not shown is that the radius of the dark matter core tends to decrease with increasing fermion mass.  If ordinary matter has a relatively large central pressure, then the small dark matter core has little effect, as seen by the red region squeezing close to the black curve in \ref{fig:MR mneq1}(d).  On the other hand, if ordinary matter has a relatively small central pressure (which, in the absence of dark matter, would be located on the tail end of the black curve with large radii), the dark matter core pulls this ordinary matter in to smaller radii, giving the low mass solutions that can be seen at the very bottom of Fig.\ \ref{fig:MR mneq1}(d).  As the fermion mass is increased further, one continues to find these low mass solutions with ever smaller masses and radii around 10 km.  Eventually solutions with planet-like masses are found, which have been dubbed ``dark compact planets" in \cite{Tolos:2015qra}.

\begin{figure}
\centering
\includegraphics[width=2.25in]{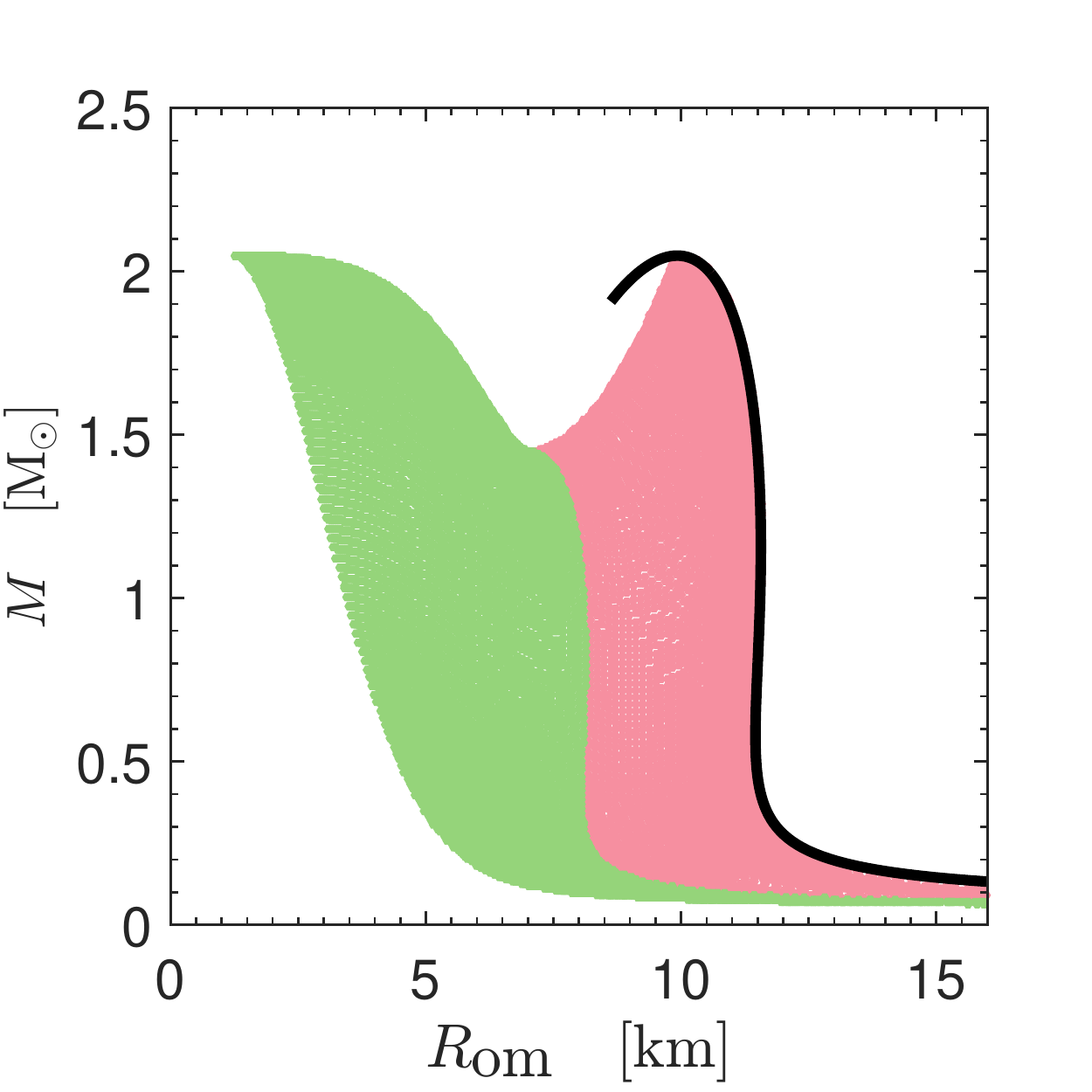}
\caption{The same as Fig.\ \ref{fig:MR m1}, except for the stable static solutions shown in Fig.\ \ref{fig:stability mirror} (dark matter is mirror dark matter).}
\label{fig:MR mir}
\end{figure}

In Fig.\ \ref{fig:MR mir} we show the MR diagram for mirror dark matter.  Compared to the free Fermi gas with $m_f = 1$ GeV in Fig.\ \ref{fig:MR m1}, the region with a dark matter core is similar.  A notable difference is an expanded region with a dark matter halo, in which we find 1--2 M$_\odot$ solutions, but with visible radii in the 2--8 km range.


\section{Radial oscillations}
\label{sec:radial}

Chandrasekhar initiated the study of stellar oscillations of neutron stars when he derived a pulsation equation whose solution gives the squared radial oscillation frequency \cite{Chandrasekhar:1964zz}.  His pulsation equation was subsequently rewritten in various ways \cite{Misner:1974qy, Chanmugam, Gondek:1997fd, Kokkotas:2000up}, in some cases to facilitate numerical solutions.  As mentioned in Sec.\ \ref{sec:stability}, the squared radial oscillation frequency can be used to determine the stability of a static solution.  Although radial oscillation modes do not couple to gravitational waves, they are, in principle, observable by the emission of electromagnetic radiation from the surface of the star (see, for example, \cite{Brillante:2014lwa}).  The hope is that their study can reveal details of the inner structure of the star.  Radial oscillation frequencies have been computed for a large number of equations of state (see, for example, \cite{Glass1983, Benvenuto1991, Gondek:1997fd, Kokkotas:2000up, VasquezFlores:2010eq, Brillante:2014lwa, DiClemente:2020szl}). 

Chandrasekhar derived his pulsation equation for a single-fluid system.  Recently, a system of pulsation equations was derived for an arbitrary number of perfect fluids with only gravitational inter-fluid interactions \cite{Kain:2020zjs} (see also \cite{Comer:1999rs}).  In \cite{Kain:2020zjs}, these equations were used to compute radial oscillation frequencies in one-, two-, and three-fluid systems, where all fluids were taken to be a free Fermi gas.  In this section we use these equations to study radial oscillations of dark matter admixed neutron stars.  This is the first time the pulsation equations of \cite{Kain:2020zjs} have been applied to a system where one of the fluids is described with a realistic equation of state.  The pulsation equations and how they are solved is reviewed in the Appendix.

Radial oscillations of dark matter admixed neutron stars is understudied.  As far as we are aware, the only works that have computed such oscillation frequencies using two-fluid methods are \cite{Leung:2011zz, Leung:2012vea, Leung:2013pra}, using the equations of \cite{Comer:1999rs}, and only \cite{Leung:2012vea} presented results beyond those used in determining stability.  (References \cite{Panotopoulos:2017eig, Panotopoulos:2018ipq} computed radial oscillation frequencies of two-fluid systems, but did so using Chandrasekhar's single-fluid pulsation equation.)  Frequencies have also been computed by Fourier transforming results from simulations using full numerical relativity \cite{ValdezAlvarado:2012xc}.  Our aim in this section is to make a systematic computation of oscillation frequencies for the fundamental radial mode.

\begin{figure}
\centering
\includegraphics[width=2.5in]{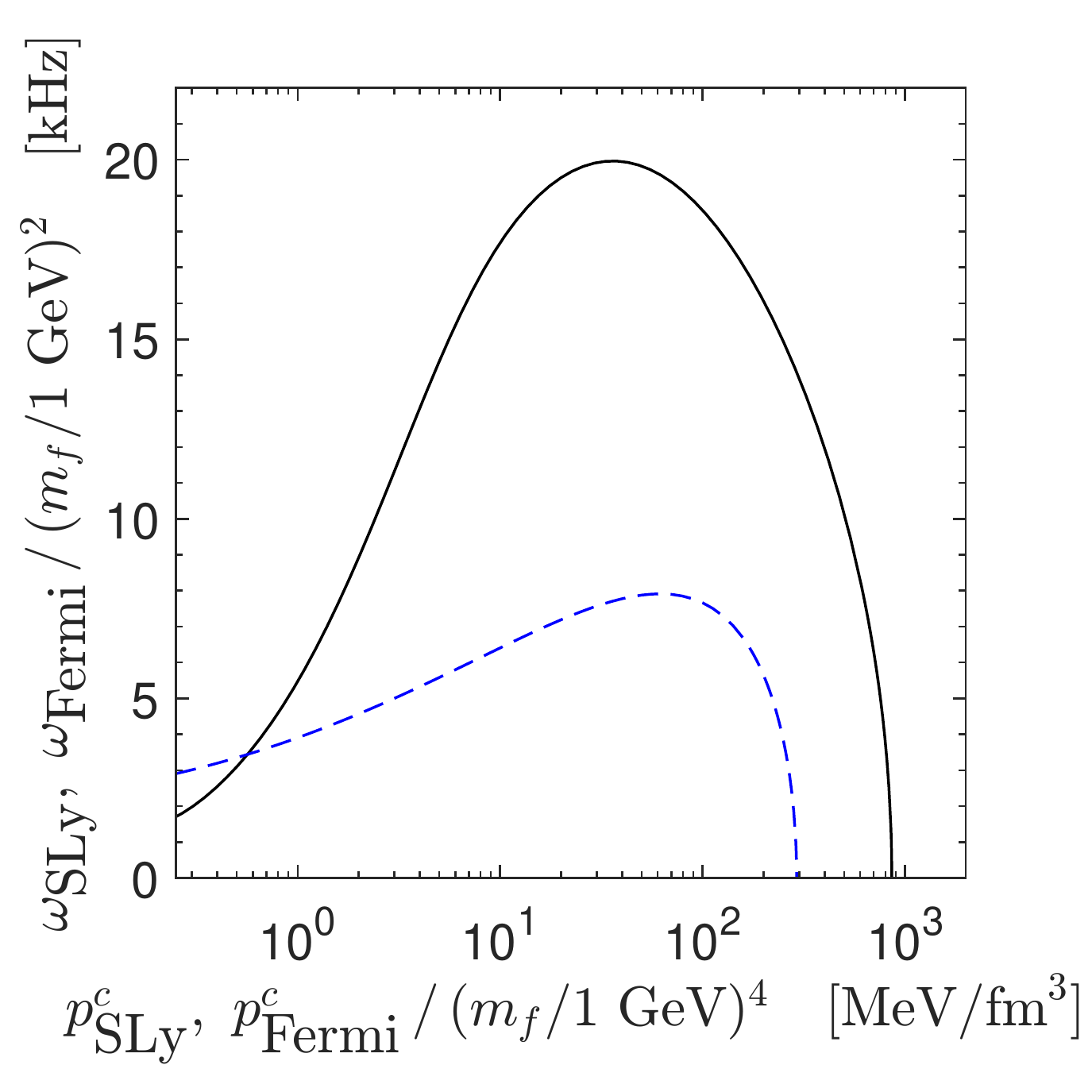}
\caption{The fundamental radial oscillation frequency, $\omega$, is plotted as a function of the central pressure for single-fluid stars using the SLy (solid black curve) and free Fermi gas (dashed blue curve) equations of state.  The curves are seen to hit zero at the critical central pressures given in Eq.\ (\ref{critical points}).  For SLy, the maximum frequency is $\omega = 19.96$ kHz, which occurs for a central pressure of $p^c = 36.02$ MeV/fm$^3$.  For the free Fermi gas, the maximum frequency is $\omega = 7.91(m_f/\text{1 GeV})^2$ kHz, which occurs for a central pressure of $p^c = 61.70(m_f/\text{1 GeV})^4$ MeV/fm$^3$.}
\label{fig:rad 1fluid}
\end{figure}

We begin first by solving Chandrasekhar's \textit{single}-fluid pulsation equation for the radial oscillation frequency, $\omega$, of the fundamental solution for single-fluid stars using the SLy and free Fermi gas equations of state.  The results are shown in Fig.\ \ref{fig:rad 1fluid} as a function of the central pressure, which uniquely identifies the static solution.  The solid black line is for SLy and the dashed blue line is for the free Fermi gas.  Note that the frequencies hit zero at the critical central pressures given in (\ref{critical points}).  This is expected, since the critical central pressures mark the point at which the static solutions transition from stable to unstable, which occurs when the (squared) radial oscillation frequency (of the fundamental mode) transitions from positive to negative.  Note also that the frequencies of a free Fermi gas with fermion mass $m_f = 1$ GeV are smaller than those for the neutron star (with the SLy equation of state) over much of the parameter space.  For $m_f = 0.5$ GeV, the dashed blue curve drops by a factor of 4, making it much smaller than the neutron star, while for $m_f = 2$ GeV, the dashed blue curve grows by a factor of 4, making it larger than the neutron star over much of the parameter space.  It is useful to keep these facts in mind in the following.

\begin{figure*}
\centering
\includegraphics[width=6.5in]{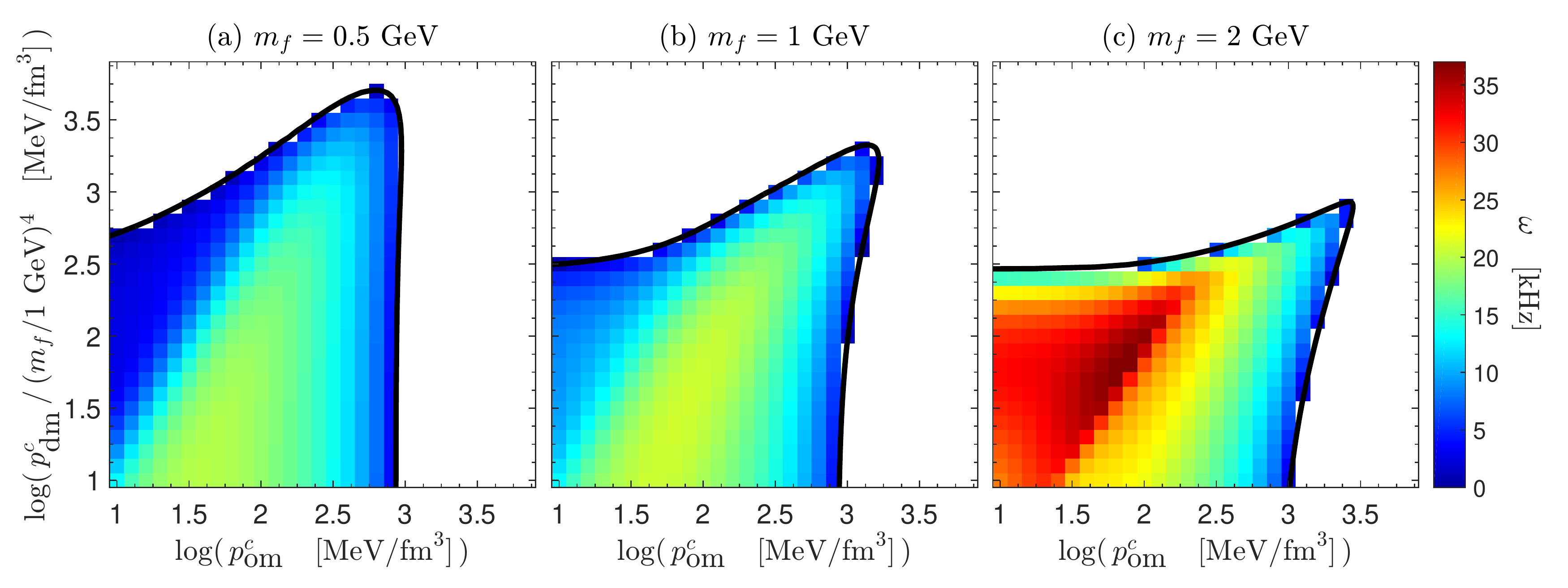}
\caption{The fundamental radial oscillation frequency, $\omega$, is plotted as a function of the central pressures $p^c_\text{om}$ and $p^c_\text{dm}$, where dark matter is taken to be a free Fermi gas with fermion masses, $m_f$, as indicated above each plot.  The thick black lines are critical curves, first shown in Figs.\ \ref{fig:stability m1} and \ref{fig:stability mneq1}.}
\label{fig:rad asym}
\end{figure*}

Unfortunately, the computation of the radial oscillation frequency in a two-fluid system is time consuming.  For this reason, we do not present results for all cases, nor over the entirety of the parameter space, that we considered previously.  In Fig.\ \ref{fig:rad asym} we show results for a free Fermi gas with fermion masses $m_f = 0.5$, 1, and 2 GeV.  In each plot, the thick black line is the critical curve first shown in Figs.\ \ref{fig:stability m1} and \ref{fig:stability mneq1}.  We can see that near the critical curves, the oscillation frequencies head toward zero, as expected.  This offers some visual evidence that the two methods for determining stability \cite{Kain:2020zjs, Henriques:1990xg} discussed in Sec.\ \ref{sec:stability} agree, though we note that we have confirmed that they agree to much higher precision than that shown in Fig.\ \ref{fig:rad asym}.

Consider now the bottom of the plot in Fig.\ \ref{fig:rad asym}(a) for $m_f = 0.5$ GeV.  The frequencies match well the single-fluid frequencies given by the solid black line in Fig.\ \ref{fig:rad 1fluid}.  This tells us that dark matter has a very small effect on the oscillation frequency in this region of parameter space.  As we move up from the bottom in Fig.\ \ref{fig:rad asym}(a), we do not see much change in frequency, and nearly no change in frequency for larger values of $p^c_\text{om}$.  This tells us that only when we have larger values of $p^c_\text{dm}$ and smaller values of $p^c_\text{om}$ is dark matter able to affect substantially the frequency.  Once we compare this to the other plots in Fig.\ \ref{fig:rad asym}, we conclude that this is because of the smaller fermion mass of $m_f = 0.5$ GeV.  We also gain insight as to why the critical curve extends past the single-fluid value of $(p_\text{dm}^c)_\text{crit}$, but not $(p_\text{om}^c)_\text{crit}$, first noticed in Sec.\ \ref{sec:stability}.  With the small fermion mass of $m_f = 0.5$ GeV, it is only after increasing $p^c_\text{dm}$ substantially is dark matter able to affect the system and cause it to be unstable.  Even then, for large $p^c_\text{om}$, the ordinary matter is always dominating and we do not find stability past the single-fluid value of $(p_\text{om}^c)_\text{crit}$.

Now consider Fig.\ \ref{fig:rad asym}(b), where we see some, though not significant, changes compared to \ref{fig:rad asym}(a).  That the changes are not significant is expected, because for both $m_f = 0.5$ and 1 GeV, the single-fluid frequencies for a free Fermi gas, as given by the dashed blue curve in Fig.\ \ref{fig:rad 1fluid}, are smaller than the single-fluid frequencies for ordinary matter, as given by the solid black curve in Fig.\ \ref{fig:rad 1fluid}.  Still, we can see in Fig.\ \ref{fig:rad asym}(b) that with the larger fermion mass, dark matter has a bigger influence over the frequency.  This connects with the fact that $p_\text{dm}^c$ does not have to be raised as high before it makes the system go unstable and that for large $p_\text{dm}^c$, we can push $p_\text{om}^c$ past its single-fluid critical value and still have stable solutions.

We do see significant changes in Fig.\ \ref{fig:rad asym}(c).  This is expected, since for $m_f = 2$ GeV the single-fluid frequency for a free Fermi gas can be larger than the single-fluid frequency for ordinary matter in Fig.\ \ref{fig:rad 1fluid}.  The larger frequencies in Fig.\ \ref{fig:rad asym}(c) coming from dark matter, in a sense, ``collide" with the smaller frequencies coming from normal matter in the center of the figure.  Interestingly, this causes the frequency to increase, as can be seen by the darker region near the center, to values larger than the maximum possible single-fluid frequencies.

\begin{figure}
\centering
\includegraphics[width=2.5in]{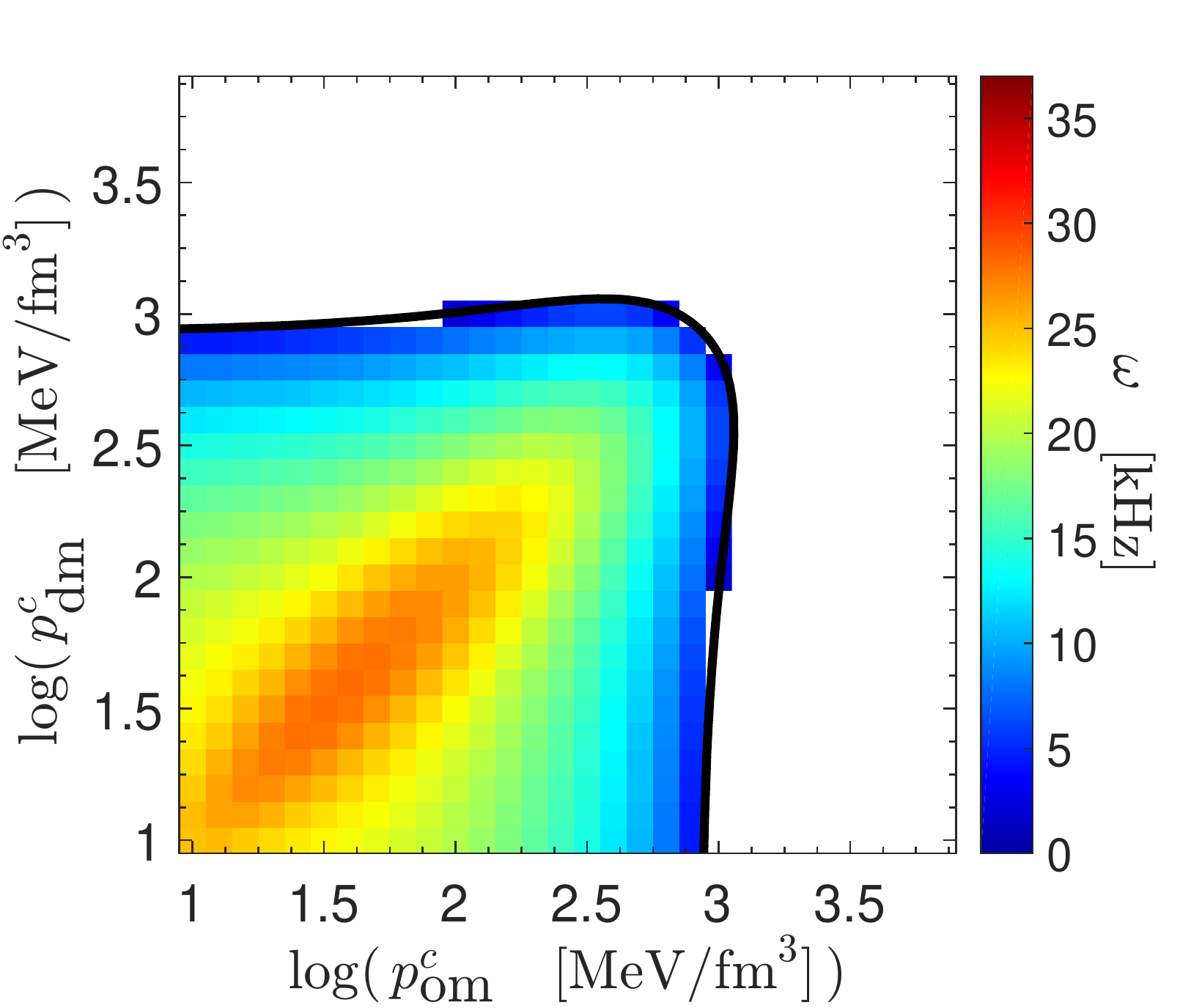}
\caption{The same as Fig.\ \ref{fig:rad asym}, except for mirror dark matter, in which dark matter has the same equation of state as ordinary matter.  The thick black line is the critical curve, first shown in Fig.\ \ref{fig:stability mirror}.}
\label{fig:rad mir}
\end{figure}

In Fig.\ \ref{fig:rad mir} we show results for mirror dark matter.  The thick black line is the critical curve first shown in Fig.\ \ref{fig:stability mirror}.  The figure is symmetric in parameter space, since the same equation of state is used for dark matter as is used for ordinary matter.  Similar to Fig.\ \ref{fig:rad asym}(c), we find a ``collision" in the center of the plot with a frequency that is larger than the maximum single-fluid frequency given by the solid black line in Fig.\ \ref{fig:rad 1fluid}.


\section{Conclusion}
\label{sec:conclusion}

We studied dark matter admixed neutron stars, which are two-fluid systems with only gravitational inter-fluid interactions.  The first fluid describes  ordinary nuclear matter and the second fluid describes dark matter.  We considered two possibilities for dark matter:\ a free Fermi gas and mirror dark matter.  Static solutions were found by solving the two-fluid TOV equations.

Our study focused on three computations.  The first was the stability of static solutions with respect to small perturbations.  Rigorous determinations of stability over large swaths of parameter space were lacking in the literature.  We presented two different ways to determine stability and computed critical curves, which separate stable solutions from unstable ones in parameter space.   Interestingly, we found stable regions of parameter space for which a naive analysis of the individual equations of state would not have deemed stable.

The second computation was for mass-radius relations from static solutions.  As an alternative to what is commonly presented in the literature, we gave mass-radius diagrams over the whole of the stable static parameter space, highlighting when dark matter acts as a core or as a halo in the star.

The third computation was the radial oscillation frequency.  As with stability, computations of the radial oscillation frequency over large swaths of parameter space were lacking in the literature.  Interestingly, our results showed that the frequencies of dark matter admixed neutron stars could be larger than the maximum possible frequencies of single-fluid stars made with the individual equations of state.


\appendix 
\section{Radial oscillations and stability}
\label{sec:appendix}

In Sec.\ \ref{sec:stability}, we discussed two methods for computing the critical curve, which separates stable static solutions from unstable ones in parameter space.  In Sec.\ \ref{sec:radial}, we computed radial oscillation frequencies.  In this appendix we review the equations used in these computations.


\subsection{Radial oscillations}

The method we use to solve for the radial oscillation frequencies was derived in \cite{Kain:2020zjs}.  We briefly review the equations and how they are solved here and refer the reader to \cite{Kain:2020zjs} for details.  To make a time-dependent perturbation around a static solution, we must first write the energy-momentum tensor, $T^{\mu\nu} = \sum_i T_i^{\mu\nu}$, in terms of the full perfect fluid form,
\begin{equation}
T_i^{\mu\nu} = (\epsilon_i + p_i) u_i^\mu u_i^\nu + p_i g^{\mu\nu},
\end{equation}
where $u^\mu_i$ is the four-velocity of the fluid, and not in terms of the static form, as was done in Eq.\ (\ref{static perfect fluid}).  Spherical symmetry sets $u^\theta = u^\phi = 0$ and we define $v_i \equiv e^{\nu/2}u_i^r$, where $\nu$ is the metric function in (\ref{metric}) (but now with a time dependence).  We can then write the metric functions, energy densities, and pressures as perturbations about their static solutions and then write the Einstein field equations and equations of motion to first order in the perturbations.  Note that $v_i$ is at the level of a perturbation, since it vanishes in the static limit.

Defining the quantity $\xi_i$ through $\partial_t \xi_i \equiv v_i$,
the perturbations are all taken to be of harmonic form,
\begin{equation} \label{perturbation}
\xi_i(t,r) = \xi_i(r) e^{i\omega t},
\end{equation} 
which defines the radial oscillation frequency, $\omega$.  We further define
\begin{equation} 
\zeta_i(r) \equiv r^2 e^{-\nu_0(r)/2} \xi_i(r),
\end{equation}
where a subscripted 0 in this section refers to a static solution.  The idea is to combine the Einstein field equations and equations of motion such that we obtain a system of pulsation equations which depend on $\zeta_i$ and its derivatives, and not on any other perturbations.

Before presenting the pulsation equations, we rewrite the metric in (\ref{metric}) as
\begin{equation}
ds^2 = - H(t,r) \sigma^2(t,r) dt^2 + \frac{dr^2}{H(t,r)} + r^2d\Omega^2,
\end{equation}
where
\begin{equation}
H(t,r) \equiv 1 - \frac{2 m(t,r)}{r}, \quad
\sigma(t,r) \equiv \frac{e^{\nu(t,r)/2}}{\sqrt{H(t,r)}},
\end{equation}
which is better suited for numerical solutions.  The equation for  the static metric function $\sigma_0(r)$ is
\begin{equation}
\frac{d\sigma_0}{dr} = \frac{4\pi  r \sigma_0}{H_0} ( \epsilon_0 + p_0),
\end{equation}
which follows from the Einstein field equations and is one of the TOV equations, but was not listed in (\ref{TOV eqs}) because, for static solutions, it decouples and is not needed.

The system of pulsation equations is \cite{Kain:2020zjs}
\begin{widetext}
\begin{align}
&\partial_r (\widehat{\Pi} \zeta_i') +(\widehat{Q}_i + \hat{\omega}^2 W_i) \hat{\zeta}_i
+ \widehat{R}
\left[ 
\left(\frac{\epsilon_{i0} + p_{i0} }{r} - p_{i0}' \right) 
\sum_j
(\epsilon_{j0} + p_{j0}) \hat{\zeta}_j
+
\frac{r^2 (\epsilon_{i0} + p_{i0})}{\hat{\sigma}_0^2 H_0}
\sum_j
\eta_j 
\right]
\notag \\
&\qquad = \widehat{S}_i \sum_j (\epsilon_{j0} + p_{j0}) \left( \hat{\zeta}_j - \hat{\zeta}_i \right)
+ 
\frac{r^2}{\hat{\sigma}_0^2 H_0}  
\widehat{R}^2
(\epsilon_{i0} + p_{i0}) 
\sum_j
\sum_k
 p_{j0} \gamma_j (\epsilon_{k0} + p_{k0}) \left( \hat{\zeta}_k - \hat{\zeta}_j \right)
\notag \\
&\qquad\qquad  +
\widehat{R} \gamma_i p_{i0}
\sum_j \left[(\epsilon_{j0}' + p_{j0}') \left( \hat{\zeta}_j - \hat{\zeta}_i \right)
+ (\epsilon_{j0} + p_{j0}) \left( \hat{\zeta}_j' - \hat{\zeta}_i' \right) \right],
\label{pulsation}
\end{align}
where a prime denotes an $r$ derivative, where $p_{i0}'$ is given by the TOV equation in (\ref{TOV eqs}), and where
\begingroup
\allowdisplaybreaks
\begin{align}
\widehat{\Pi}_i &= \frac{1}{r^2}p_{i0} \gamma_i \hat{\sigma}^2_0 H_0
\notag \\
W_i &= \frac{1}{r^2 H_0} (\epsilon_{i0} + p_{i0} )
\notag \\
\widehat{Q}_i &=
-\frac{ \hat{\sigma}^2_0 H_0}{r^2}
\biggl\{
\frac{3}{r} p_{i0}'
+ 
 \biggl[
\frac{8\pi}{H_0} p_0
(\epsilon_{i0} + p_{i0}) 
+ \left(\frac{4\pi r}{H_0} \epsilon_0 - \frac{m_0}{r^2 H_0} \right)  \left( \frac{\epsilon_{i0} + p_{i0} }{r} - p_{i0}'\right)
\biggr] 
\biggr\}
\notag \\
\widehat{R} &= 4\pi 
\frac{\hat{\sigma}_0^2}{r}
\notag \\
\widehat{S}_i &= \widehat{R}
\biggr\{
(\gamma_i-1)
p_{i0}' 
+ \gamma_i' p_{i0} +
\gamma_i  p_{i0} \bigg[ \frac{8\pi r}{H_0} (\epsilon_0 + p_0 )  - \frac{1}{r}\biggr]
\biggr\}
\notag \\
\gamma_i &= \left(1 + \frac{\epsilon_{i0}}{p_{i0}}\right) \frac{\partial p_{i0}}{\partial \epsilon_{i0}}.
\label{pulsation defs}
\end{align}
\endgroup
\end{widetext}
Those quantities with a hat have been scaled by powers of $\sigma^c_0$, the central value of $\sigma_0$.  This has the effect of changing the boundary conditions and making the equations easier to solve.  Note that the right hand side of Eq.\ (\ref{pulsation}) vanishes for a single fluid, in which case the left hand side is equivalent to Chandrasekhar's pulsation equation \cite{Chandrasekhar:1964zz}.  Though equivalent, the left hand side of Eq.\ (\ref{pulsation}) is not written in an identical form to Chandrasekhar's because, in the presence of multiple fluids, terms cannot cancel and combine in the same way.

To actually solve the pulsation equations, we define $\eta_i \equiv \widehat{\Pi}_i \hat{\zeta}_i'$ and solve the system of first order differential equations made up of Eq.\ (\ref{pulsation}), but written in terms of $\eta_i$, and 
\begin{equation} \label{pulsation 2}
\hat{\zeta}_i'  = \frac{\eta_i}{\widehat{\Pi}_i}.
\end{equation}
In the two-fluid case, the inner boundary conditions are $m = \zeta_{i} = 0$ and $\hat{\sigma}_0 = \eta_1 = 1$.  The outer boundary conditions for the static variables are as discussed in Sec.\ \ref{sec:equations} and for the perturbations are $\eta_i(R_i) = 0$.  The undetermined parameters are the inner boundary condition for $\eta_2$ and the value of scaled squared radial oscillation frequency $\hat{\omega}^2$.  These are determined using the shooting method.  Once a solution to the pulsation equations is found, so too are $\sigma^c_0$, which follows from the outer value of $\hat{\sigma}_0$, and $\hat{\omega}^2$.  The squared radial oscillation frequency is then given by $\omega^2 = (\sigma^c_0 \hat{\omega})^2$.


\subsection{Stability}

In Sec.\ \ref{sec:stability}, we discussed two methods for determining whether a solution to the TOV equations is stable with respect to small perturbations.  The first method is to compute the squared radial oscillation frequency using the methods of the previous subsection.  Importantly, one must compute oscillation frequencies for the \textit{fundamental} solution, which has the smallest frequency, and not for excited solutions.  In practice, this is easily done by making sure that the $\eta_i$ do not have nodes, i.e.\ values for which they equal zero before the edge of their respective fluid.   If $\omega^2<0$ for the fundamental solution, then the perturbation in Eq.\ (\ref{perturbation}) describes a growing mode and the corresponding static solution is unstable; if $\omega^2>0$, the corresponding static solution is stable.  It is possible for the squared frequency of the fundamental solution to be negative while the excited frequencies are positive, which is why it is the fundamental frequency that must be computed.  In this paper, including in Sec.\ \ref{sec:radial}, we only compute radial oscillation frequencies for fundamental solutions. 

The transition from stable to unstable for static solutions occurs at those points in parameter space where $\omega^2 = 0$.  Such points map out the critical curves shown in Figs.\ \ref{fig:stability m1}--\ref{fig:stability mirror}.  In this way, the computation of the squared radial oscillation frequency using the methods of the previous subsection can be used to compute critical curves.

The second method we use for computing the critical curve is from \cite{Henriques:1990xg}.  The idea is straightforward.  We still wish to find those points in parameter space where $\omega = 0$.  From  Eq.\ (\ref{perturbation}) we find that, in this case, the perturbation is time independent and thus the perturbed static solution is itself a static solution.  This is convenient, since it means we need only deal with static solutions.  Further, the perturbations (including those with time-dependence) preserve the total fluid number, $N_i$, and total mass, $M$, of the system.  Putting these facts together, if there is a point in the parameter space of static solutions where $\omega = 0$, then there must be some direction in parameter space, given by vector $\mathbf{p}$, that preserves $\omega = 0$ as well as the total fluid numbers and mass of the system.  This immediately gives Eq.\ (\ref{stability conditions}), which is what we used to compute the critical curves in Sec.\ \ref{sec:stability}.




%

\end{document}